\documentclass[a4paper]{jpconf}
\usepackage{graphicx}
\begin{document}
\title{Photon-photon production of lepton, quark and meson pairs
in peripheral heavy ion collisions}

\author{Antoni Szczurek}

\address{Institute of Nuclear Physics PAN, Krak\'ow, Poland\\
         Rzesz\'ow Univeristy, Rzesz\'ow, Poland}

\ead{antoni.szczurek@ifj.edu.pl}

\author{Mariola K{\l}usek-Gawenda}

\address{Institute of Nuclear Physics PAN, Krak\'ow Poland}

\begin{abstract}
We review our recent results on exclusive production of
$\mu^+ \mu^-$, heavy quark-antiquark, and meson-antimeson pairs
in ultraperipheral, ultrarelativistic heavy ion collisions.
\end{abstract}

\section{Introduction}

Ultrarelativistic collisions of heavy ions provide a nice oportunity
to study photon-photon collisions \cite{review}.
One can expect an enhancement of the cross section for the reactions of
this type compared to proton-proton or $e^+ e^-$ collisions which
is caused by a large charges of the colliding ions. In this type of
reactions virtual (almost real) photons couple to the nucleus as a whole.
Naively the enhancement of the cross section is proportional to 
$Z_1^2 Z_2^2$ which is a huge factor. We have discussed recently that 
the inclusion of realistic charge distributions and realistic nucleus 
charge form factor
makes the cross section smaller than the naive predictions.
Many processes has been discussed in the literature.
Recently we have also studied some of them.

We have discussed production of $\mu^+ \mu^-$ pairs \cite{KS_mumu}
heavy-quark heavy-antiquark pairs \cite{KSMS_qqbar}
as well as production of two mesons: $\rho^0 \rho^0$ 
pairs \cite{KSS_rhorho}, $\pi^+ \pi^-$ pairs \cite{KS_pipi}
as well as of $D \bar{D}$ meson pairs \cite{LS_DDbar}.

Here we shall summarize the recent works.

\section{Formalism}

\subsection{Equivalent Photon Approximation}

\begin{figure}[!h]    
\begin{center}
\includegraphics[width=0.3\textwidth]{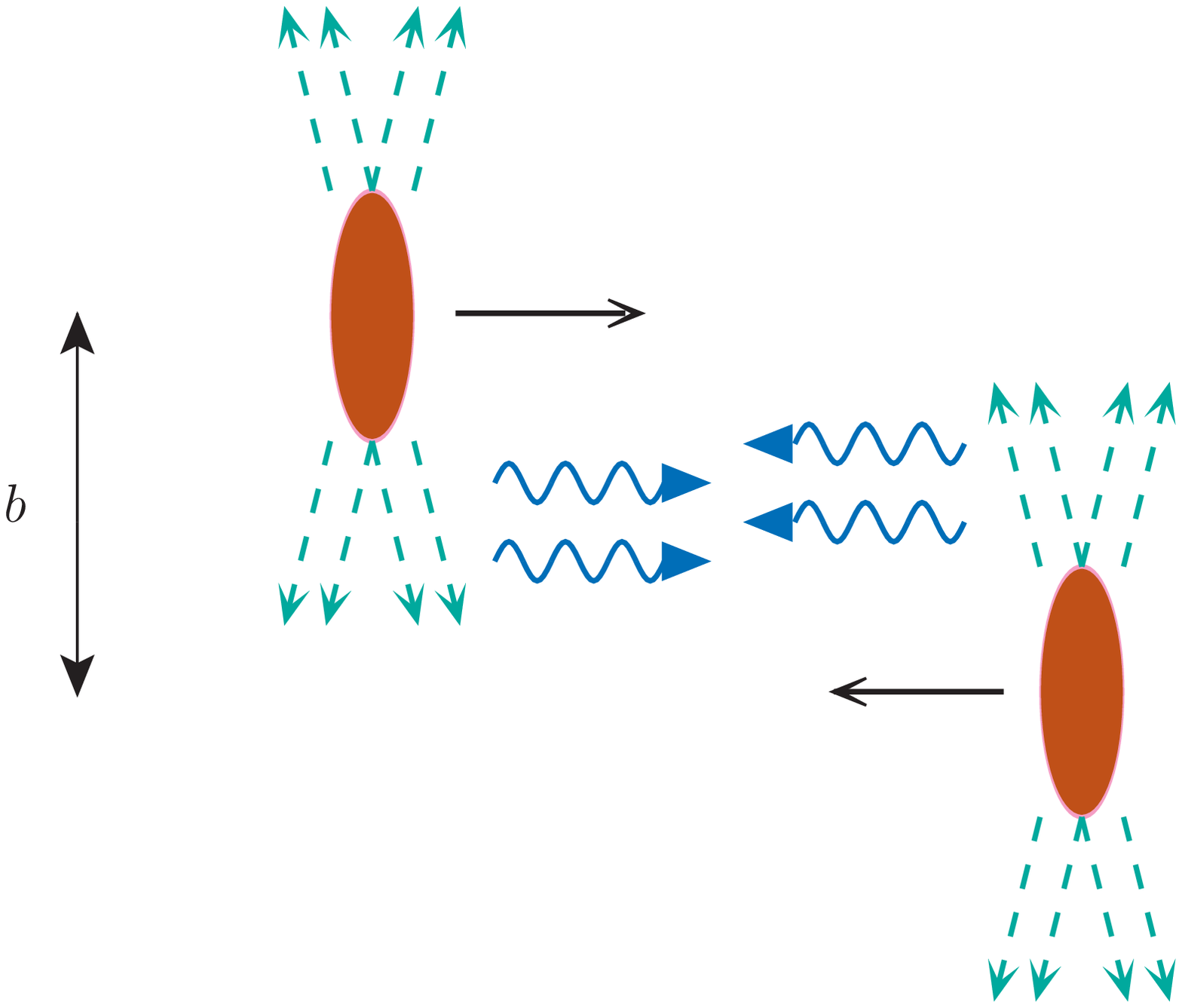}
\includegraphics[width=0.3\textwidth]{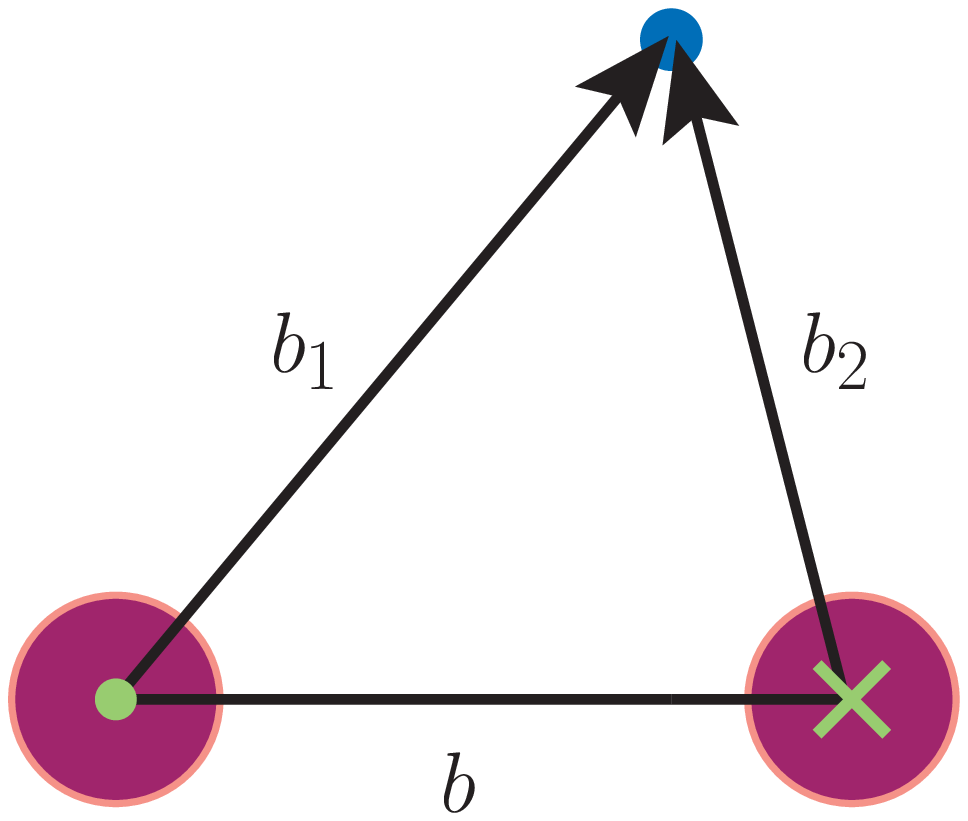}
\end{center}
   \caption{\label{fig:EPA}
   \small
A schematic picture of the collision and
the quantities used in the impact parameter calculation.
}
\end{figure}

The equivalent photon approximation is a standard semi--classical 
alternative to the Feynman rules for calculating cross sections 
of electromagnetic interactions \cite{Jackson}. 
This picture is illustrated in Fig.~\ref{fig:EPA} where one can 
see a fast moving nucleus with the charge $Ze$. 
Due to the coherent action of all protons in the nucleus, 
the electromagnetic field surrounding (the dashed lines are lines 
of electric force for a particles in motion) 
the ions is very strong. This field can be viewed as a cloud of 
virtual photons. 
In the collision of two ions, these quasireal photons can collide with 
each other and with the other nucleus. The strong 
electromagnetic field is a source of photons that can induce 
electromagnetic reactions on the second ion. 
We consider very peripheral collisions i.e. we assume 
that the distance between nuclei is bigger 
than the sum of radii of the two nuclei. 
Fig.~\ref{fig:EPA} explains also the quantities 
used in the impact parameter calculation. In the right panel we can see 
a view in the plane perpendicular to the direction of motion of 
the two ions.
In order to calculate the cross section of a process it is 
convenient to introduce the following kinematic variables:
\begin{itemize}
\item $x = \omega/E_A$, where $\omega$ energy of the photon 
and the energy of the nucleus 
\item $E_A=\gamma A m_{proton} = \gamma M_A$ where $M_A$ is 
the mass of the nucleus and $E_A$ is the energy of the nucleus
\end{itemize}

Below we consider a generic reaction $AA \to AA c_1 c_2$ and
later consider different examples when $c_1$ and $c_2$ are leptons,
quarks or mesons.
In the equivalent photon approximation the total cross section is 
calculated by the convolution:
\begin{equation}
\sigma\left(AA \rightarrow c_1 c_2 AA ;s_{AA}\right) =  
 \int   {\hat \sigma} \left(\gamma\gamma\rightarrow c_1 c_2;
W_{\gamma \gamma} = \sqrt{x_1 x_2 s_{AA}} \right) 
 {\rm d}n_{\gamma\gamma}\left(x_1,x_2,{\bf b}\right)  .
\end{equation}
%
%

%

%
%
The luminosity function $d n_{\gamma \gamma}$ above can be expressed 
in term of flux factors of photons prescribed to each of the nucleus:
\begin{equation}
{\rm d}n_{\gamma\gamma}\left(\omega_1,\omega_2,{\bf b}\right) = 
\int  S^2_{abs}\left( {\bf b} \right)
{\rm d^2}{\bf b}_1 N \left(\omega_1,{\bf b}_1 \right)  {\rm d^2}{\bf b}_2  N\left(\omega_2,{\bf b}_2 \right) \frac{{\rm d}\omega_1}{\omega_1} \frac{{\rm d}\omega_2}{\omega_2}.
\end{equation}

The presence of the absorption factor $S^2_{abs}\left({\bf b} \right)$ 
assures that we consider only peripheral collisions, when the nuclei 
do not touch each other i.e. do not undergo nuclear breakup. 
In the first approximation this can be taken into account by the
following approximation:  
\begin{equation}
 S^2_{abs}\left({\bf b} \right)=\theta \left({\bf b}-2R_A \right) = \theta \left(|{\bf b}_1-{\bf b}_2|-2R_A \right) \; .
\end{equation}

In the present case, we concentrate on processes with final 
nuclei in the ground state. 
The electric field force can be expressed through the charge 
form factor of the nucleus \cite{KS_mumu}.

The total cross section for the $AA \rightarrow c_1 c_2 AA $ 
process can be factorized into an equivalent photons spectra
and the $\gamma \gamma \to c_1 c_2$ subprocess cross section as:
\begin{eqnarray}
&& \sigma\left(AA \rightarrow c_1 c_2 AA ; s_{AA}\right)  = 
\int  
{\hat \sigma}\left(\gamma\gamma\rightarrow c_1 c_2; 
\sqrt{4 \omega_1 \omega_2}  \right) \, 
\theta \left(|{\bf b}_1-{\bf b}_2|-2R_A \right)  
\nonumber \\
&& N\left(\omega_1,{\bf b}_1 \right) N\left(\omega_2,{\bf b}_2 \right)
   {\rm d^2}{\bf b}_1 
 {\rm d^2}{\bf b}_2   \frac{{\rm d}\omega_1}{\omega_1} \frac{{\rm d}\omega_2}{\omega_2} \; .
\label{eq.tot_cross_section}
\end{eqnarray}
%
%

We introduce the invariant mass of the $\gamma \gamma $ system: 
$W_{\gamma \gamma}=\sqrt{4 \omega_1 \omega_2}$. 
Additionally, we define 
$ Y=\frac{1}{2} \left( y_{c_1}+y_{c_2}\right)$ rapidity of 
the outgoing $c_1 c_2$ system. 
Making the following transformations:
\begin{equation}
\omega_1 = \frac{W_{\gamma \gamma}}{2}e^Y, \qquad \omega_2 = \frac{W_{\gamma \gamma}}{2}e^{-Y} \; ,
\label{eq:omega}
\end{equation}
\begin{equation}
\frac {{\rm d}\omega_1} {\omega_1} \frac{{\rm d}\omega_2} {\omega_2} = \frac{2}{W_{\gamma \gamma}} {\rm d}W_{\gamma \gamma} {\rm d} Y \; ,
\label{eq:transf}
\end{equation}
\begin{equation}
{\rm d} \omega_1 {\rm d} \omega_2 \to {\rm d} W_{\gamma \gamma} {\rm d}Y
, \quad
\left|\frac{\partial \left( \omega_1, \omega_2 \right) }{\partial \left( W_{\gamma \gamma}, Y \right)}  \right| = \frac{W_{\gamma \gamma }}{2}
\; ,
\label{eq:transf_jac}
\end{equation}
formula (\ref{eq.tot_cross_section}) can be written in an 
equivalent way as:
\begin{eqnarray}
&& \sigma\left(AA \rightarrow  c_1 c_2 AA ; s_{AA}\right)  = 
 \int   {\hat \sigma}\left(\gamma\gamma\rightarrow c_1 c_2;
   W_{\gamma \gamma}  \right) \theta \left(|{\bf b}_1-{\bf b}_2|-2R_A \right)  
\nonumber \\
&&  N\left(\omega_1,{\bf b}_1 \right) 
 N\left(\omega_2,{\bf b}_2 \right) 
  \times  {\rm d^2}{\bf b}_1  {\rm d^2}{\bf b}_2   
\frac{W_{\gamma \gamma}}{2} {\rm d}W_{\gamma \gamma} {\rm d} Y
\; .
\label{eq.tot_cross_section_WY}
\end{eqnarray}
Finally the cross section can be expressed as the five-fold 
integral:
\begin{eqnarray}
&& \sigma\left(AA \rightarrow  c_1 c_2 AA ; s_{AA}\right)  = 
 \int  {\hat \sigma}\left(\gamma\gamma\rightarrow \mu^+ \mu^-;
  W_{\gamma \gamma}  \right) \theta \left(|{\bf b}_1-{\bf b}_2|-2R_A
\right)  
\nonumber \\
&&  \times   N \left(\omega_1,{\bf b}_1 \right) N\left(\omega_2,{\bf
    b}_2 \right) 
 2 \pi b_m \, {\rm d} b_m \, {\rm d} b_x \, {\rm d} b_y \frac{W_{\gamma \gamma}}{2} {\rm d}W_{\gamma \gamma} {\rm d} Y \, , 
\label{eq.tot_cross_section_our}
\end{eqnarray}
where $\vec{b}_x \equiv (b_{1x}+b_{2x})/2$,
      $\vec{b}_y \equiv (b_{1y}+b_{2y})/2$ and
$\vec{b}_m = \vec{b}_1 - \vec{b}_2$ have been introduced.
The formula above is used to calculate the total cross section
for the $A A \to A A c_1 c_2$ reaction as well as distributions
in $b = b_m$, $W_{\gamma \gamma} = M_{c_1 c_2}$ and
$Y(c_1 c_2)$.

Different forms of charge form factors of nucleus were used in 
the literature. 
We compare the equivalent photon spectra 
for realistic charge distribution and for 
the case of monopole form factor. 
A compact formula how the photon flux depends on
the charge form factors can be found in \cite{review}. 
\begin{equation}
N \left( \omega,b \right) = \frac{Z^2 \alpha_{em}}{\pi^2} 
\frac{1}{b^2 \omega} 
\left( \int u^2 J_1 \left( u \right) F \left( \sqrt{\frac{ \left(\frac{b \omega}{\gamma} \right)^2 +u^2}{b^2}} \right) 
 \frac{1}{ \left( \frac{b \omega}{\gamma} \right)^2 + u^2} du \right)^2,
\label{basic_EPA_flux}
\end{equation}
where $J_1$ is the Bessel function of the first kind and $q$ is 
momentum of the quasireal photon. 
The calculations with the help of realistic form factor are rather 
laborious, so often a simpler formula with monopole form factor
is used \cite{HTB94}. 

%
%

\subsection{Charge form factor of nuclei}

The charge distribution in nuclei is usually obtained from
elastic scattering of electrons from nuclei \cite{Barrett_Jackson}.
The charge distribution obtained from those experiments is 
often parametrized with the help of two--parameter Fermi model 
\cite{nuclear_density}:
\begin{equation}
 \rho \left( r \right) = \rho_0 \left( 1 + \exp \left( \frac{r-c}{a} \right) \right)^{-1},
\end{equation}
where $c$ is the radius of the nucleus, $a$ is the so-called diffiusness parameter of the charge density. 

\begin{figure}[!h]   
\begin{center}
\includegraphics[width=0.4\textwidth]{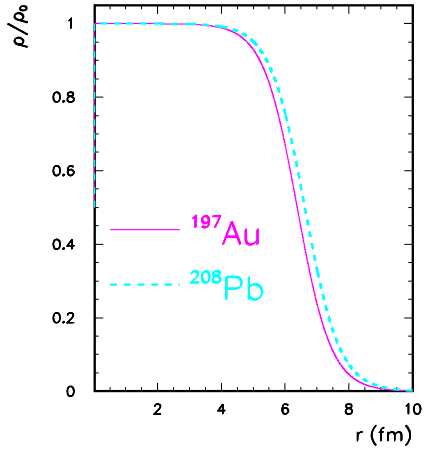}
\end{center}
   \caption{\label{fig:density}
   \small
The ratio of $\rho$ the charge distibution to $\rho_0$,
the density in the center of nucleus.
}
\end{figure}
Fig.~\ref{fig:density} shows the charge density normalized to 
unity. The correct normalization is:
$\rho_{ 0,\, Au}(0) = \frac{0.1694}{A} fm^{-3}$ for Au and 
$\rho_{ 0,\, Pb}(0) = \frac{0.1604}{A} fm^{-3}$ for Pb. 

\begin{figure}[!h]
\begin{center}   
\includegraphics[width=0.4\textwidth]{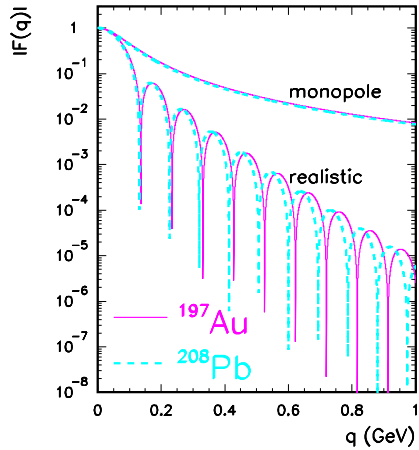}
\end{center}
    \caption{\label{fig:ff_all}
    \small
The moduli of the charge form factor $F_{em} \left ( q \right)$ of the $^{197}Au$ and $^{208}Pb$ nuclei for realistic charge distributions. For comparison we show the monopole form factor for the same nuclei.
}
\end{figure}

Mathematically the charge form factor is the Fourier transform of 
the charge distribution \cite{Barrett_Jackson}:
\begin{equation}
 F(q) = \int \frac{4 \pi}{q} \rho \left( r \right) \sin \left( qr \right)
 r dr 
\; .
\label{Fourier}
\end{equation}
Fig.~\ref{fig:ff_all} shows the moduli of the form factor calculated
from Eq.(\ref{Fourier}) as a 
function of momentum transfer. 
Here one can see many oscillations characteristic for relatively 
sharp edge of the nucleus. We show results for the gold 
(solid line) and lead (dashed line) nuclei for realistic charge 
distribution. For comparison we show the monopole form factor 
often used in the literature. The two form factors coincide only in 
a very limited range of $q$.

The monopole form factor \cite{HTB94}:
\begin{equation}
 F(q^2) = \frac{\Lambda^2}{\Lambda^2 + q^2}.
\end{equation}
leads to a simplification of many formulae for production of pairs of
particles via photon-photon subprocess in nucleus-nucleus collisions.
In our calculation  $\Lambda$ is adjusted to reproduce 
root mean square radius $\Lambda = \sqrt{\frac{6}{<r^2>}}$
with the help of experimental data \cite{nuclear_density}.

\subsection{Exclusive production of $\mu^+ \mu^-$ pairs}

Elementary cross section for charged leptons can be calculated
within Quantum Electrodynamics. Several groups have made relevant 
calculations (see e.g. \cite{BB87,HTB99,Serbo,Baltz} and references therein).

Recently we have performed calculation of exclusive production of $\mu^+ \mu^-$
and explored potential of RHIC and LHC in this respect.
In Ref.\cite{KS_mumu} we have presented several distributions in
muon rapidity and transverse momentum for RHIC and LHC experiments,
including experimental acceptances. We have demonstrated how important
is inclusion of realistic form factor in order to obtain realistic
distributions of muons for RHIC and LHC. 
Many previous calculations in the literature concentrated rather on 
the total cross section and did not pay attention
to differential distributions. Hovever, future experiments will measure
the cross section in very limited part of the phase space.
Here we wish to present only some selected examples.

\begin{figure}[!h]    
\begin{center}
\includegraphics[width=0.35\textwidth]{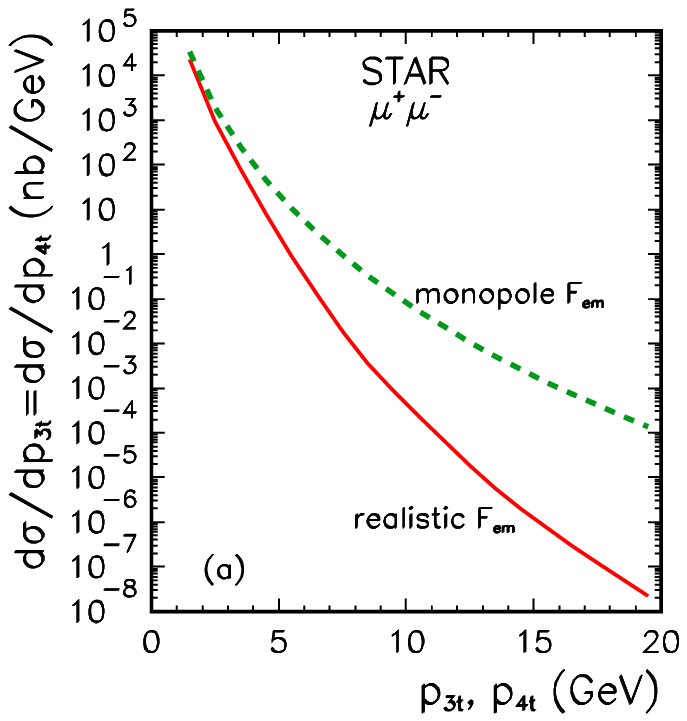}
\includegraphics[width=0.35\textwidth]{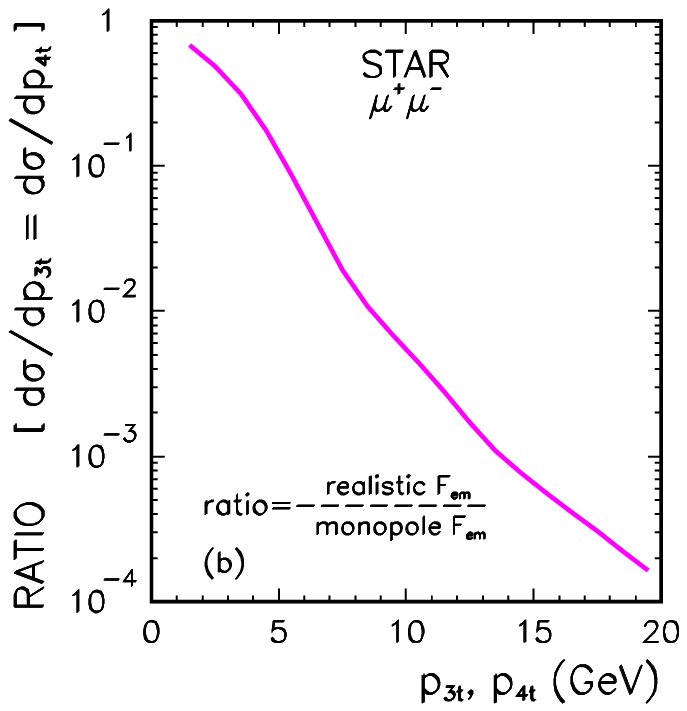}
\end{center}
   \caption{\label{fig:dsig_dp3t_STAR}
   \small 
$\frac{{\rm d} \sigma}{{\rm d} p_{3t}}$ (left) and the ratio (right)
for the STAR conditions: $y_3, y_4 \in (-1,1)$, $p_{3t}, p_{4t} \geq$
 1 GeV and $W_{NN}=200$ GeV.
}
\end{figure}

The distribution in the muon transverse momentum for STAR detector
is shown in Fig.\ref{fig:dsig_dp3t_STAR}. The STAR rapidity cuts
-1 $< y_3, y_4 <$ 1 are taken here into account. 
As can be seen from the figure, the inclusion of realistic charge 
distribution is here extremely important. The relative 
effect of damping of the cross section with respect to the results 
with the monopole charge form factor (often used in the literature) 
is shown in the right panel.
At $p_t$ = 10 GeV the damping factor is as big as 100!
Experiments at RHIC have a potential to confirm this prediction.

\begin{figure}[!h]
\begin{center}   
\includegraphics[width=0.35\textwidth]{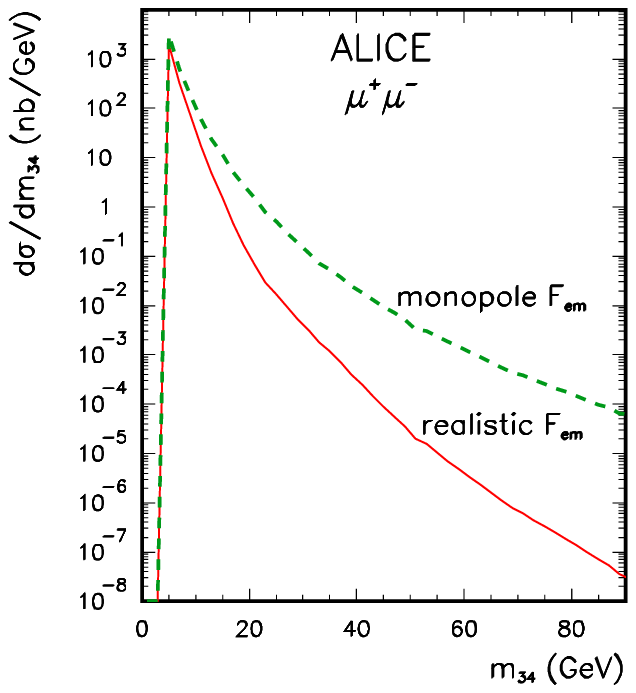}
\includegraphics[width=0.35\textwidth]{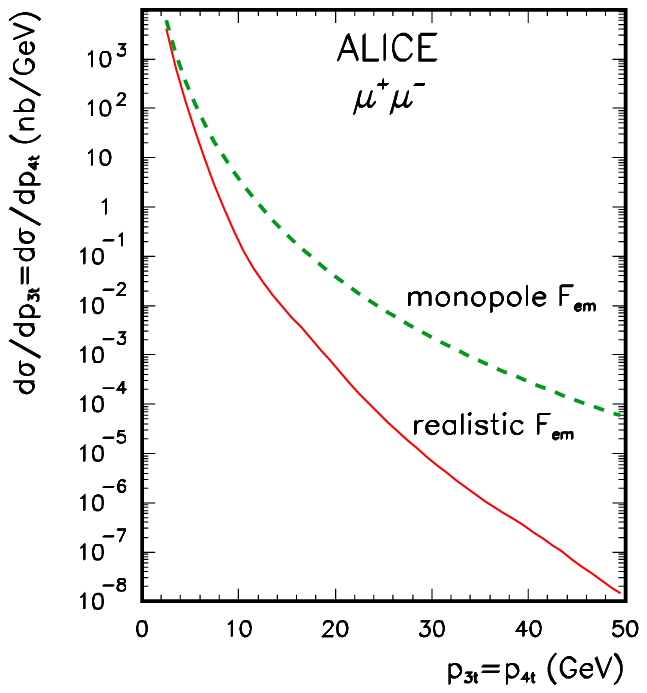}
\end{center}
   \caption{\label{fig:ALICE_varia}
   \small 
$\frac{{\rm d} \sigma}{{\rm d} W_{\gamma \gamma}}$ (left) and
$\frac{{\rm d} \sigma}{{\rm d} p_{3t}} = \frac{{\rm d} \sigma}{{\rm d} p_{4t}}$ (right)
for ALICE conditions: $y_3, y_4 = (3,4)$, $p_{3t}, p_{4t} \geq$ 2 GeV.
}
\end{figure}

The ALICE collaboration can measure only forward muons with
psudorapidity 3 $< \eta <$ 4 and has relatively low cut on muon 
transverse momentum $p_t >$ 2 GeV.
In Fig.\ref{fig:ALICE_varia} (left panel) we show invariant mass 
distribution of dimuons for monopole and realistic form factors
including the cuts of the ALICE apparatus.
The bigger invariant mass, the bigger the difference
between the two results.
The same is true for distributions in muon transverse
momenta (see the right panel). 

\subsection{Exclusive production of $c \bar c$ and $b \bar b$}

\begin{figure}[htb]
\begin{center}   
\includegraphics[width=.3\textwidth]{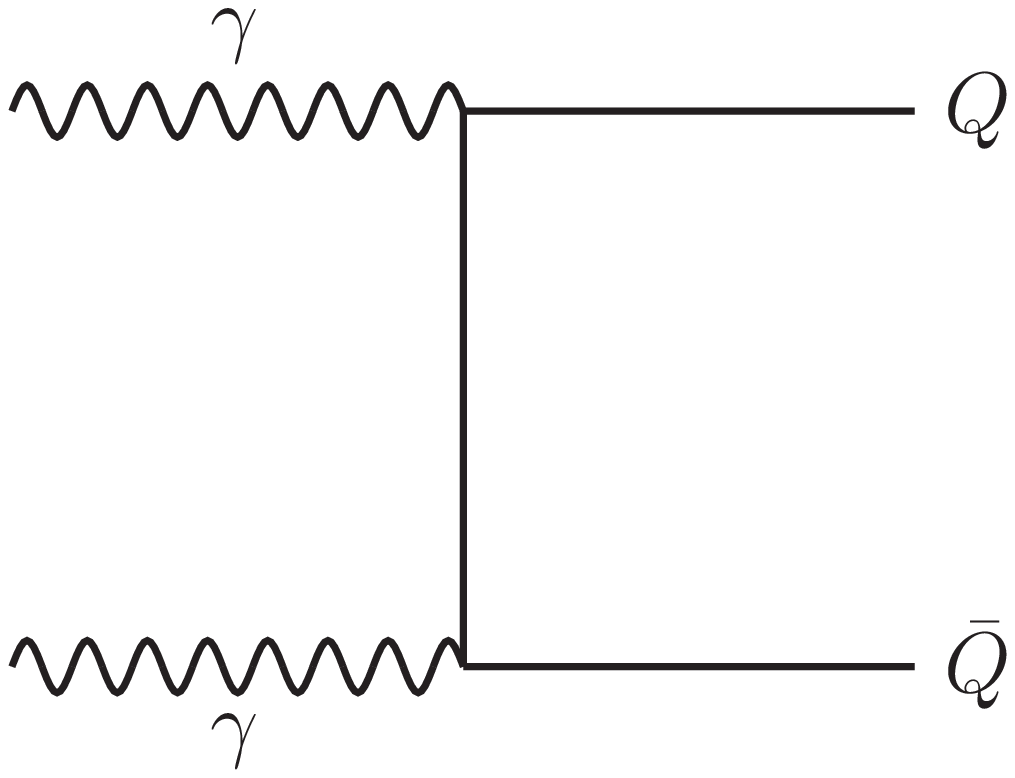}
\includegraphics[width=.3\textwidth]{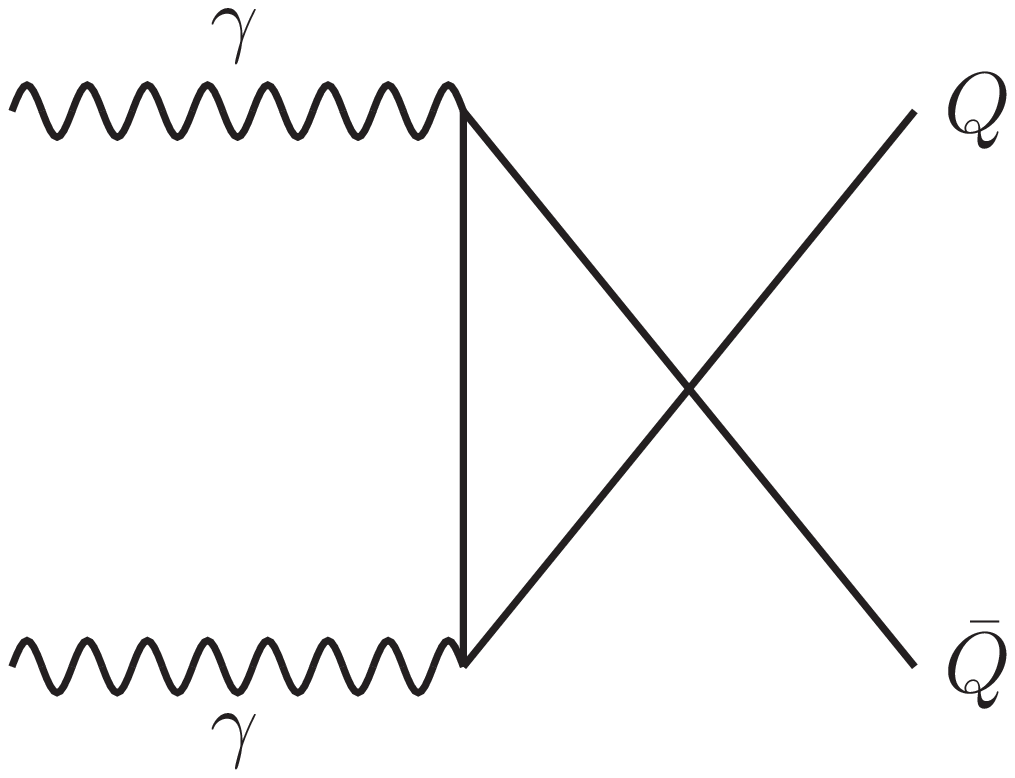}
\end{center}
   \caption{\label{fig:Born}
   \small Representative diagrams for the Born amplitudes. 
   }
\end{figure}

\begin{figure}[htb]
\begin{center}
\includegraphics[width=.5\textwidth]{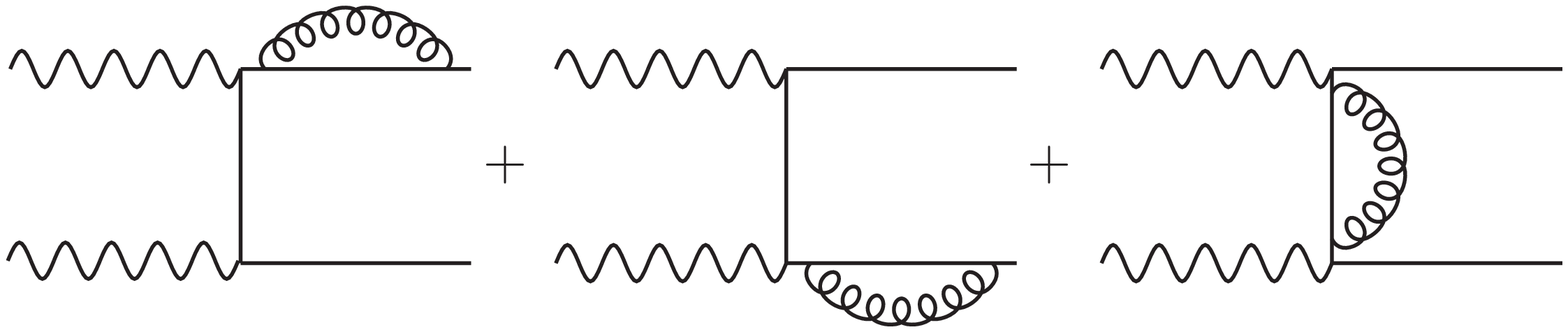}
\includegraphics[width=.5\textwidth]{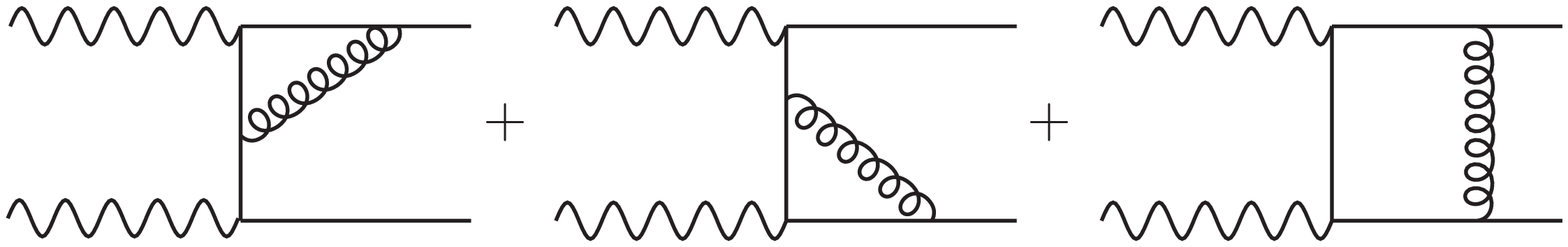}
\includegraphics[width=.5\textwidth]{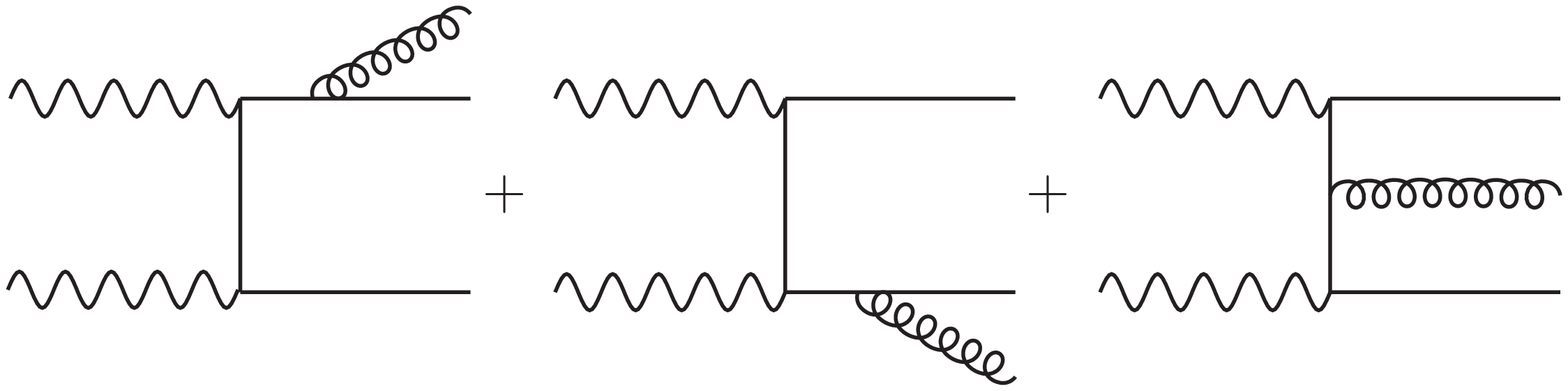}
\end{center}
   \caption{\label{fig:QCD}
   \small Representative diagrams for the leading--order QCD corrections. 
   }
\end{figure}

\begin{figure}[htb]
\begin{center}
\includegraphics[width=.3\textwidth]{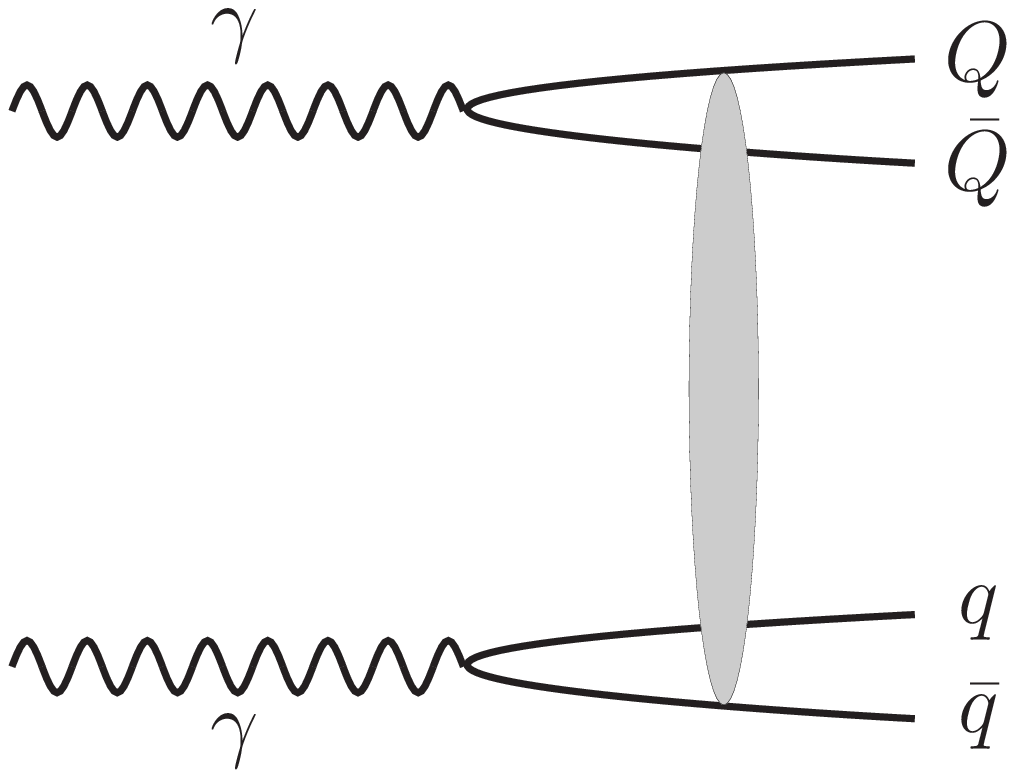}=
\includegraphics[width=.3\textwidth]{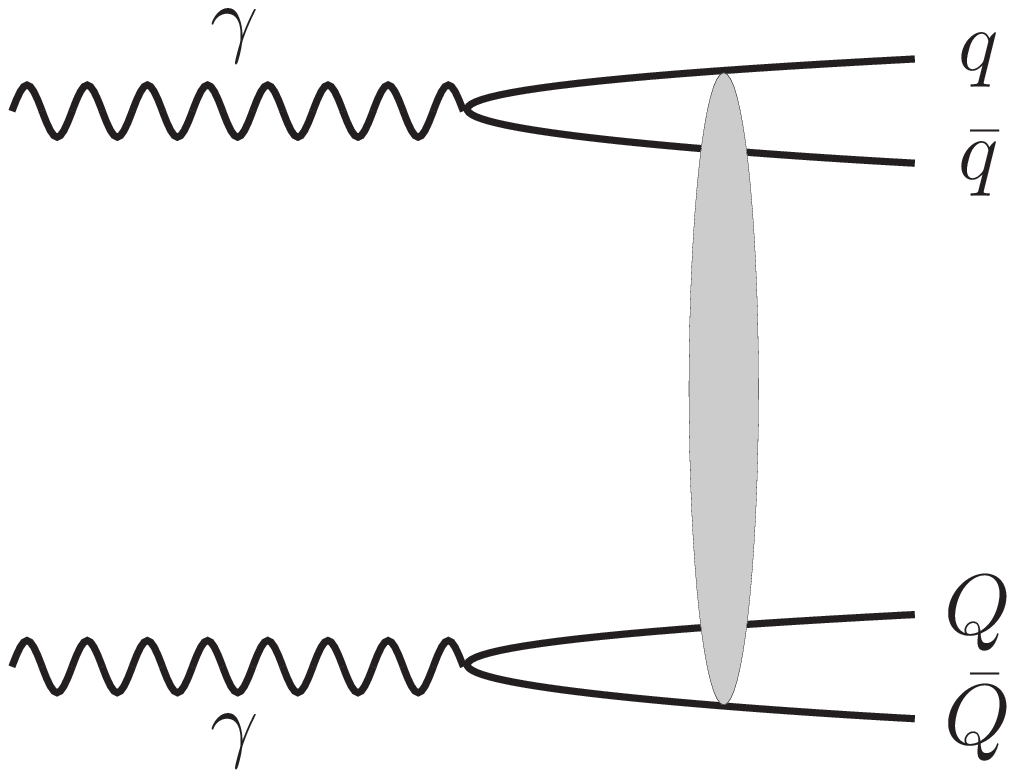}
\end{center}
   \caption{\label{fig:QQqq}
   \small Representative diagrams for $Q \bar Q q \bar q$ production. 
   The oval in the figure means a complicated interaction which is
   described here 
   in the saturation model as explained in the main text. 
   }
\end{figure}

\begin{figure}
\begin{center}
\includegraphics[width=.3\textwidth]{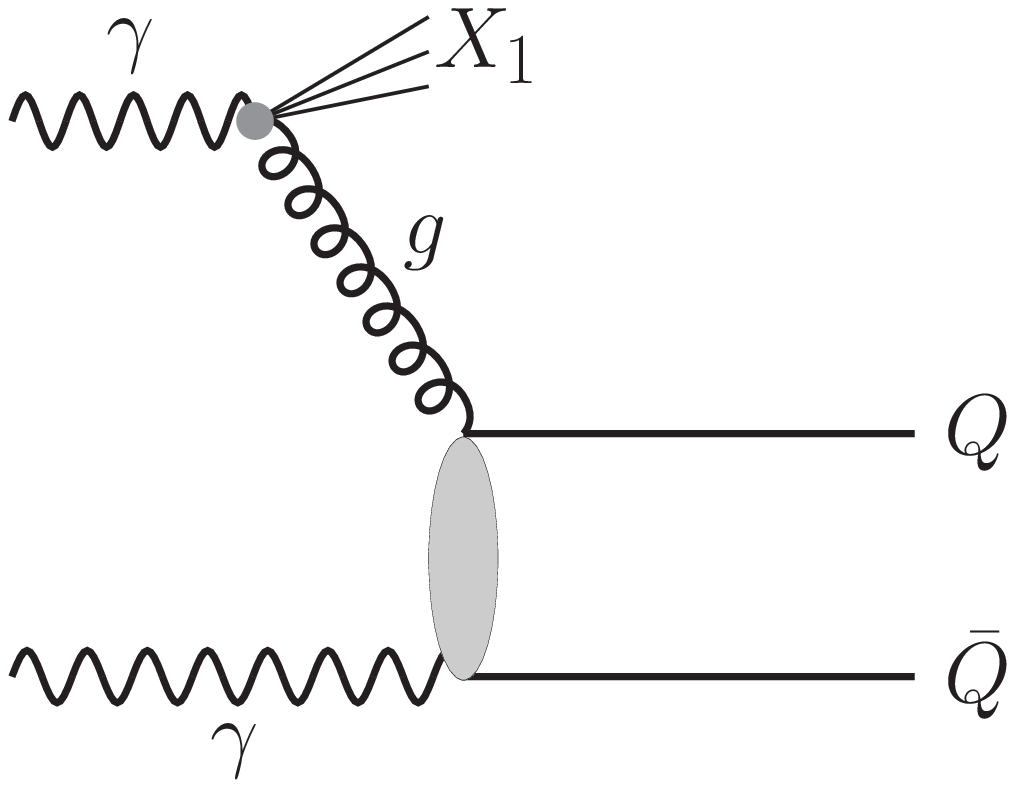}
\includegraphics[width=.3\textwidth]{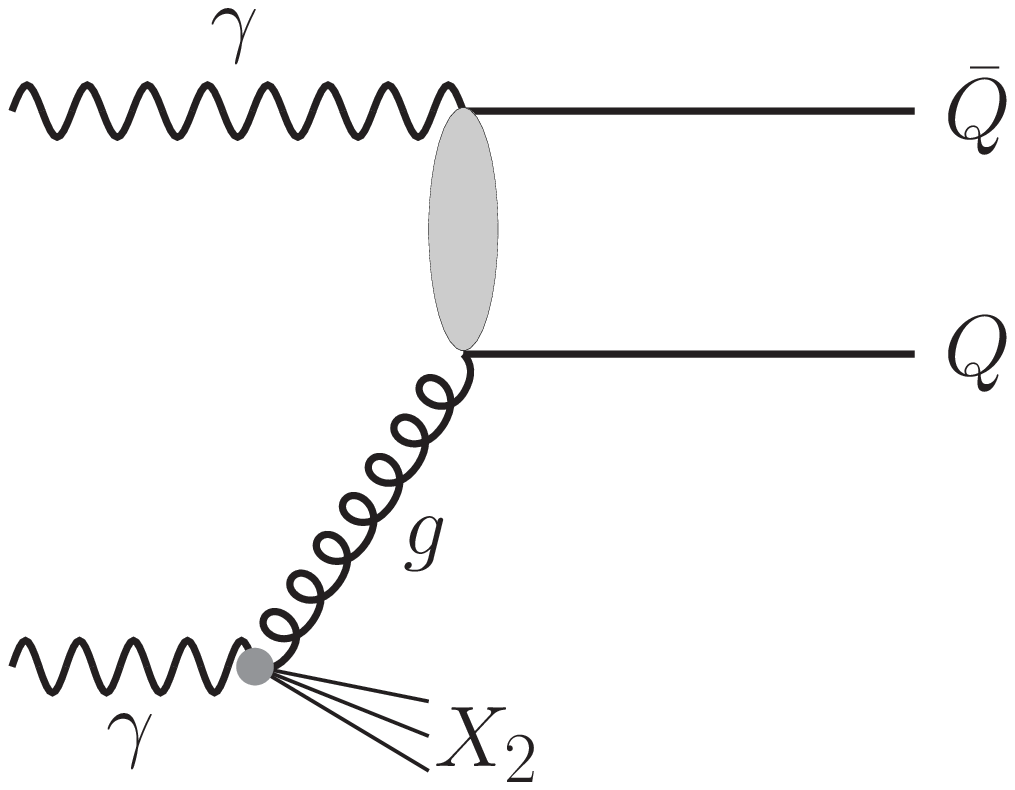}
\end{center}
   \caption{\label{fig:sr}
   \small Representative diagrams for the single-resolved mechanism.
   The shaded oval means either t- or u- diagrams
   shown in Fig. \ref{fig:Born}. 
   }
\end{figure}

In Fig.\ref{fig:Born},\ref{fig:QCD},\ref{fig:QQqq},\ref{fig:sr} 
we show several photon-photon processes leading to 
the $Q \bar Q$ in the final state.
In the following we shall discuss them one by one.


Let us start with the Born direct contribution. The leading--order
elementary cross section for $\gamma \gamma \to Q \bar{Q}$ as a function
of $W_{\gamma \gamma}$ takes a simple
form which differs from that for $\gamma \gamma \to l^+ l^-$ by color
factors and fractional charges of quarks.

%
%
In the current calculation we take the following heavy quark
masses: $m_c=1.5$ GeV, $m_b=4.75$~GeV.
It is obvious that the final $Q \bar Q$ state cannot be observed 
experimentally due to the quark confinement and rather heavy mesons have to be
observed instead. Presence of additional few light mesons is rather natural. 
This forces one to include also more complicated final states.

In contrast to QED production of lepton pairs in photon-photon
collisions, in the case of $Q \bar Q$
production one needs to include also higher-order QCD processes
which are known to be rather significant.
Here we include leading--order corrections only for the direct contribution.
In $\alpha_s$-order there occur one-gluon bremsstrahlung diagrams
($\gamma \gamma \to Q \bar Q g$) and interferences of the Born
diagram with self-energy diagrams (in $\gamma \gamma \to Q \bar Q$)
and vertex-correction diagrams (in $\gamma \gamma \to Q \bar Q$).
The relevant diagrams are shown in Fig.\ref{fig:QCD}.
We have followed the approach presented in Ref. \cite{KKMV09}.
The QCD corrections can be written as:
\begin{equation}
  \sigma_{\gamma \gamma \to Q \bar{Q} \left( g \right) }^{QCD}(W_{\gamma \gamma})= 
  N_c e_Q^4 \frac{2 \pi \alpha_{em}^2}{W_{\gamma \gamma}^2} C_F \frac{\alpha_s}{\pi} f^{\left( 1 \right)}.
\label{eq:QCD}
\end{equation}
The function $f^{\left( 1 \right)}$ is calculated using a code
provided by the authors of Ref. \cite{KKMV09}.
In the present analysis the scale of $\alpha_s$ is fixed at $\mu^2=4m_Q^2$.

We include also the subprocess $\gamma \gamma \to Q
\bar{Q} q \bar{q}$, where $q$ ($\bar q$) are $u,\, d,\, s$,
quarks (antiquarks). The cross section for this mechanism can be
easily calculated in the color dipole framework
\cite{TKM,szczurek}. In the dipole--dipole approach
\cite{szczurek} the total cross section for the $\gamma \gamma \to Q
\bar{Q}$ production can be expressed as:
\begin{eqnarray}
&&\sigma^{4q}_{\gamma \gamma \to Q \bar{Q}} \left( W_{\gamma \gamma}
\right) 
\nonumber \\
&& = \sum_{f_2 \neq Q} \int \left \vert \Phi ^{Q \bar{Q}} \left( \rho_1,z_1 \right) \right \vert^2 \left 
                                  \vert \Phi ^{f_2 \bar{f_2}} \left(
                                    \rho_2, z_2 \right) \right \vert^2
\sigma_{dd} \left( \rho_1, \rho_2, x_{Qf} \right) d^2 \rho_1 dz_1 d^2
\rho_2 dz_2 
\nonumber \\
&& + \sum_{f_1 \neq Q} \int \left \vert \Phi ^{f_1 \bar{f_1}} \left(\rho_1, z_1 \right) \right \vert^2 \left 
                                  \vert \Phi ^{Q \bar{Q}} \left( \rho_2,
                                    z_2 \right) \right \vert^2
\sigma_{dd} \left( \rho_1, \rho_2, x_{fQ} \right) d^2 \rho_1 dz_1 d^2
\rho_2 dz_2 \, ,
\label{sig_dd}
\end{eqnarray}
where $\Phi ^{Q \bar{Q}} \left( \rho,z \right)$ are the quark -- antiquark
wave functions of the photon in the mixed representation and
$\sigma_{dd}$ is the dipole--dipole cross section. Eq.(\ref{sig_dd}) is
correct at sufficiently high energy $W_{\gamma \gamma} \gg 2m_Q$.
At lower energies, the proximity of the kinematical threshold must be
taken into account.
In Ref. \cite{TKM} a phenomenological saturation--model inspired 
parametrization for the azimuthal angle averaged dipole--dipole cross
section has been proposed:
\begin{equation}
\sigma^{a,b}_{dd} = \sigma^{a,b}_0 \left[ 1- \exp \left( -
\frac{r^2_{\rm eff}}{4R^2_0 \left( x_{ab} \right)} \right)
\right].
\end{equation}
Here, the saturation radius is defined as:
\begin{equation}
R_0 \left( x_{ab} \right) = \frac{1}{Q_0} \left( \frac{x_{ab}}{x_0} \right)^{-\lambda/2}
\end{equation}
and the parameter $x_{ab}$ which controls the energy dependence is given by:
\begin{equation}
x_{ab} = \frac{4m_a^2 + 4m_b^2}{W_{\gamma \gamma}^2}.
\end{equation}
The effective radius is parametrized as
$r_{\rm eff}^2= (\rho_1\rho_2)^2/(\rho_1+\rho_2)$ \cite{TKM} .
Some other parametrizations of the dipole-dipole cross section 
were also discussed in the literature.
The cross section for the $\gamma \gamma \to Q \bar Q q \bar q$ 
process here is much bigger than the one corresponding 
to the tree-level Feynman diagram  as it effectively 
resums higher-order QCD contributions.

As discussed in Ref. \cite{szczurek} the $Q \bar Q q \bar q$ component 
have very small overlap with the single-resolved component 
because of quite different final state, 
so adding them together does not lead to double counting.
The cross section for the single-resolved contribution can be written as:
\begin{eqnarray}
&&\sigma_{1-res} \left( s \right) = 	\int d x_1 \left[g_1 \left( x_1,
    \mu^2 \right) \hat{\sigma}_{g \gamma} \left( \hat{s} = x_1 s \right)
\right] +
\nonumber \\							
&&\int d x_2 \left[g_2 \left( x_2, \mu^2 \right) \hat{\sigma}_{\gamma g} \left( \hat{s} = x_2 s \right) \right],
\end{eqnarray}
where $g_1$ and $g_2$ are gluon distributions in photon $1$ or photon $2$ 
and $\hat{\sigma}_{q \gamma}$ and $\hat{\sigma}_{\gamma g}$ are elementary cross sections. 
In our calculation we take the gluon distribution from Ref. \cite{GVR1992}.

\begin{figure}[htb]
\begin{center}                  
\includegraphics[width=0.35\textwidth]{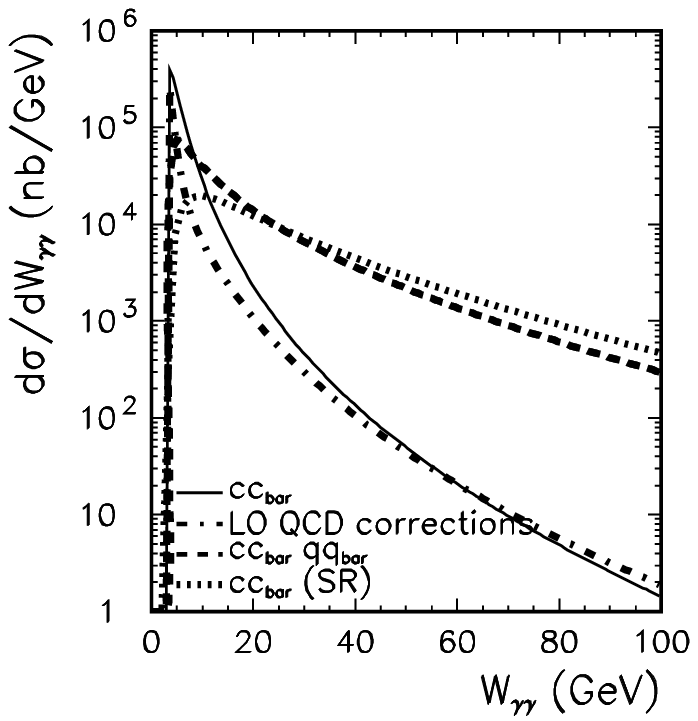}
\includegraphics[width=0.35\textwidth]{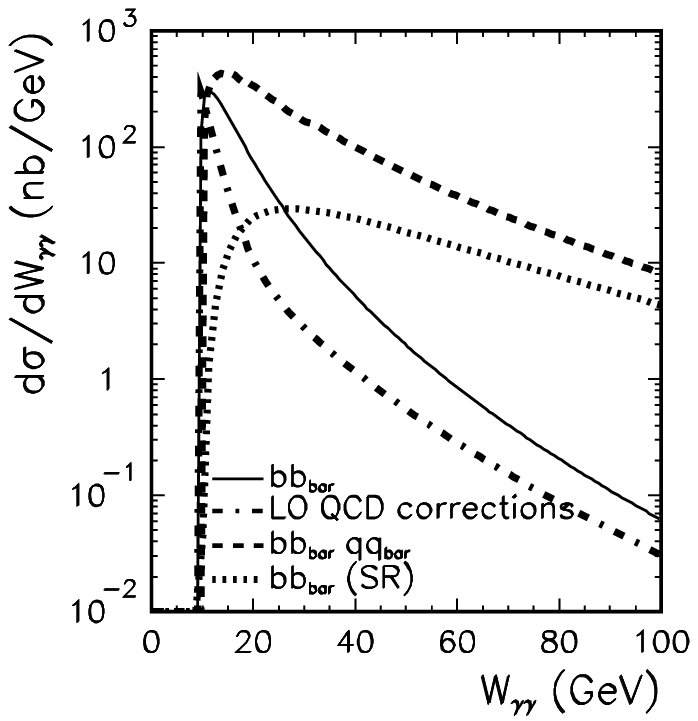}
\end{center}
   \caption{\label{fig:dsig_dw_EPA_wc}
   \small The nuclear cross section
as a function of photon--photon subsystem energy $W_{\gamma \gamma}$ 
in EPA. The solid line denotes the results corresponding to the Born
amplitude ($c\bar{c}$ -left panel and $b\bar{b}$ -right panel). 
The leading--order QCD corrections are shown by the dash-dotted line.
For comparison we show the differential
distributions in the case when an additional pair of light quarks
is produced in the final state (dashed lines) 
and for the single-resolved components (dotted line). }
\end{figure}

Elementary cross sections have been presented and discussed
in Ref.\cite{KSMS_qqbar}. Here we show only nuclear cross sections.
In Fig. \ref{fig:dsig_dw_EPA_wc} we compare the contributions of
the different mechanisms as a function of the photon--photon
subsystem energy.
For the Born case it is identical as a distribution
in quark-antiquark invariant mass.
In the other cases the photon--photon subsystem 
energy is clearly different than the $Q \bar Q$ invariant mass. 
These distributions reflect the energy dependence of the
elementary cross sections.
Please note a sizable contribution of the leading--order corrections
close to the threshold and at large energies for 
the $c \bar c$ case.
Since in this case $W_{\gamma \gamma} > M_{Q \bar Q}$, it becomes
clear that the $Q \bar Q q \bar q$ contributions must have much
steeper dependence on the $Q \bar Q$ invariant mass than the
direct one which means that large $Q \bar Q$ invariant masses are
produced mostly in the direct process. In contrast, small
invariant masses (close to the threshold) are populated dominantly
by the four--quark contribution. Therefore, measuring the
invariant mass distribution one can disentangle some of the different
mechanisms. As far as this is clear for the $c \bar c$ it is less
transparent and more complicated for the $b \bar b$ production. In
the last case the experimental decomposition may be in practice
not possible.

In Table 1 we show partial contribution of different
subprocesses discussed above.

{\tiny

\begin{table}

\begin{center}
\caption{Partial contributions of different mechanisms at $\sqrt{s_{NN}}$ = 5.5 TeV.}

\begin{tabular}{|c||c||c|c|c|c|} \hline

                   		& $\sigma_{tot}$	& Born		& QCD-corr.	& 4-q	& Sin.-res.      \\ \hline \hline

 $c \overline{c}$ 		& 2.47   $m b $     & 42.5 \% 	& 14.6 \%			& 27.1 \%	& 15.8 \%			\\ \hline
 $b \overline{b}$ 		& 10.83 $\mu b$    	& 18.9 \%   & 7.7 \%			& 64.5 \%	& 8.9 \%				\\ \hline

\end{tabular}

\end{center}

\end{table}

}

\subsection{Exclusive production of $\pi^+ \pi^-$ pairs}

\begin{figure}[htb]
\begin{center}                  
\includegraphics[width=0.2\textwidth]{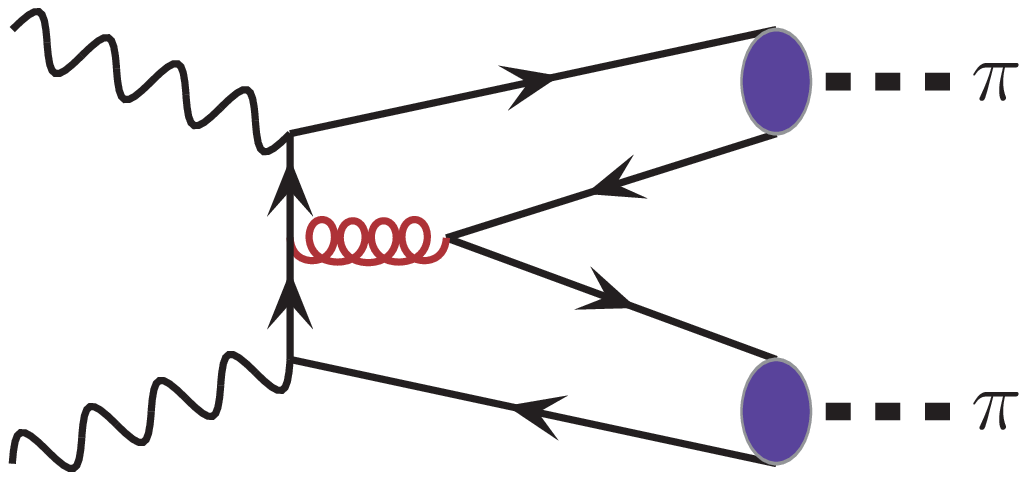}
\includegraphics[width=0.2\textwidth]{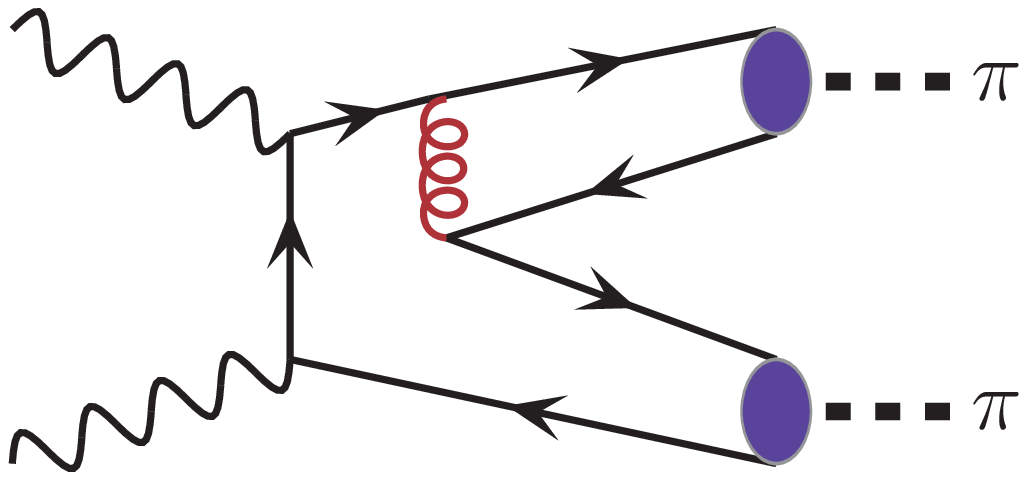}
\includegraphics[width=0.2\textwidth]{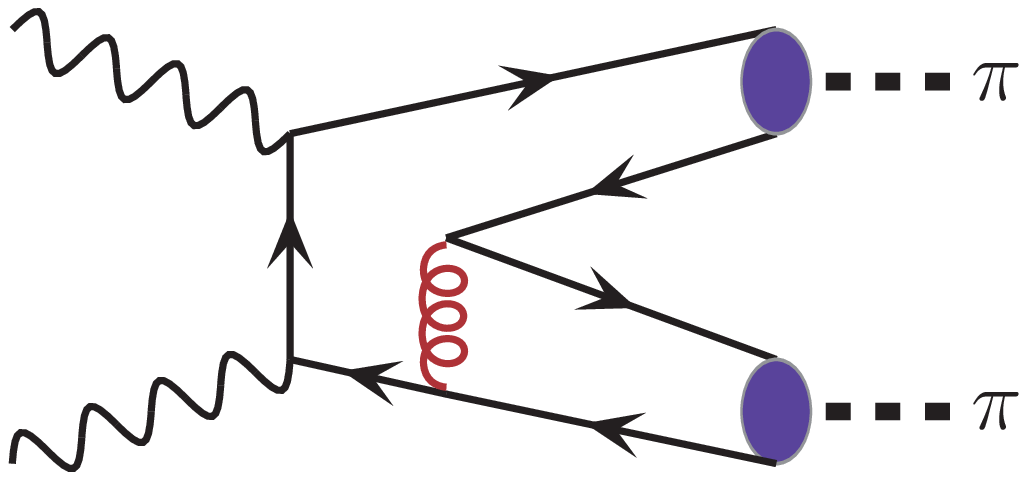}
\includegraphics[width=0.2\textwidth]{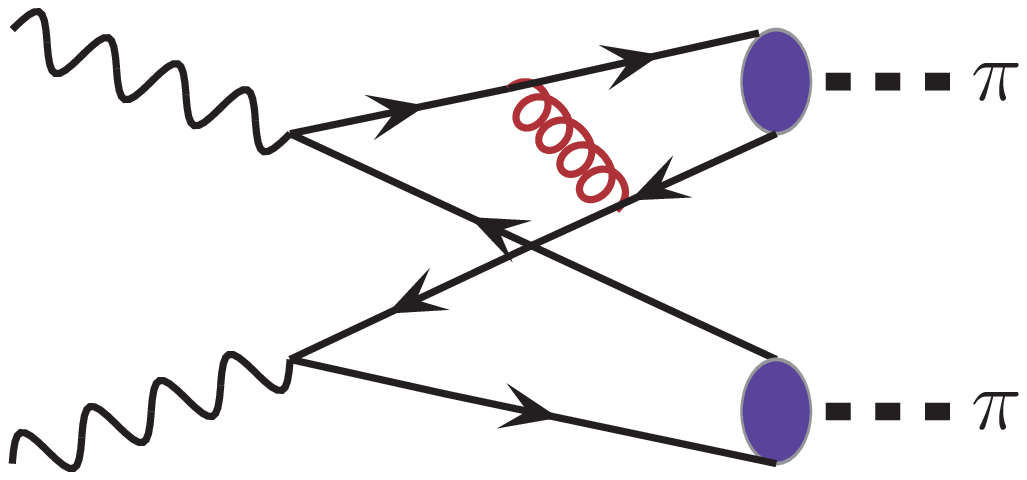}
\includegraphics[width=0.2\textwidth]{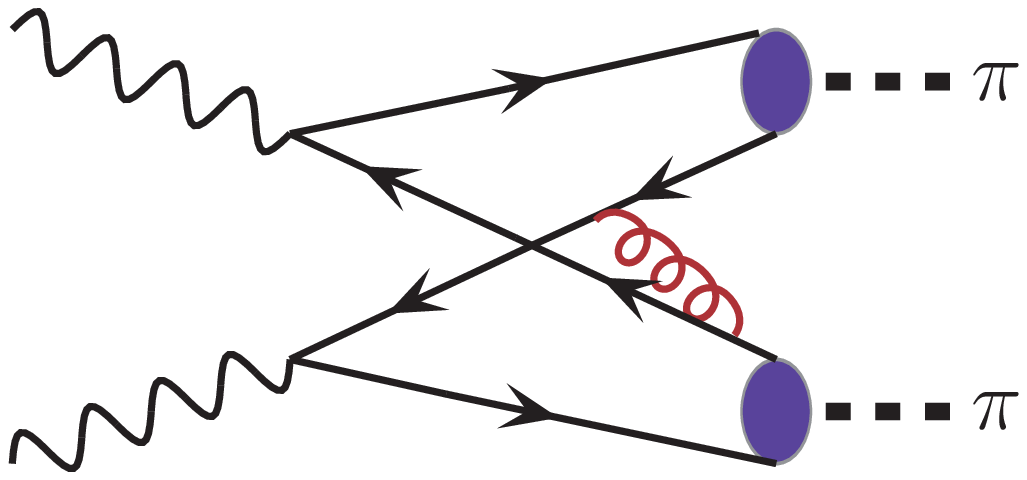}
\end{center}
   \caption{\label{fig:Feynman}\textsl{}
   \small Typical Feynman diagrams describing 
   the $\gamma \gamma \to ( q \bar{q}) (q \bar{q}) \to \pi \pi$ 
   amplitude in the LO pQCD.
}
\end{figure}

In this subsection we discuss production of only ``large''
invariant mass $\pi^+\pi^-$ pairs. Brodsky and Lepage
developed a formalism \cite{BL81} how to calculate relevant cross section.
Typical diagrams of the Brodsky-Lepage formalism are shown in 
Fig. \ref{fig:Feynman}.
The invariant amplitude for the initial helicities of two photons 
can be written as the following convolution:
\begin{eqnarray}
&& \mathcal{M}  \left( \lambda_1, \lambda_2 \right)  = 
\int_0^1 dx \int_0^1 dy \, 
\phi_\pi \left( x, \mu^2_x \right) 
 T^{\lambda_1 \lambda_2}_H \left( x,y, \mu^2 \right) \phi_\pi \left( y, \mu^2_y \right),
 \label{eq.amp}
\end{eqnarray}
where
$\mu_x = \min \left( x, 1-x  \right) \sqrt{s(1-z^2)}$,
$\mu_y = \min \left( y, 1-y  \right) \sqrt{s(1-z^2)}$; $z= \cos \theta$ 
\cite{BL81}.
We take the helicity dependent hard scattering amplitudes from Ref.~\cite{JA}.
These scattering amplitudes are different for $\pi^+ \pi^-$
and $\pi^0 \pi^0$.
The distribution amplitudes are subjected to the ERBL pQCD evolution 
\cite{BL79,ER}.
The scale dependent quark distribution amplitude of the pion
can be expanded in term of the Gegenbauer polynomials:
\begin{equation}
\phi_\pi \left( x, \mu^2 \right) = 
\frac{f_\pi}{2 \sqrt{3}}
6x \left( 1-x \right) 
\sum_{n=0}^{\infty '} C_n^{3/2} 
\left(2x-1 \right) a_n \left(\mu^2 \right)  \; .
\end{equation}
%
%
%
where $f_\pi$ is the pion decay constant. 

Different distribution amplitudes have been used in the past. 
Recently Wu and Huang \cite{WH} proposed a new distribution amplitude 
(based on a light-cone wave function):
\small{
\begin{eqnarray}
&& \phi_\pi \left( x, \mu_0^2 \right)  = 
  \frac{\sqrt{3}A \, m_q \beta}{2 \sqrt{2} \pi^{3/2} f_\pi} 
 \sqrt{x \left( 1-x \right)} 
\left( 1+B \times C_2^{3/2} \left(2x-1 \right) \right) 
\nonumber \\
&& \left( \mbox{Erf} \left[ \sqrt{\frac{m_q^2+\mu_0^2}{8 \beta^2
        x \left(1-x \right)}} \right] 
- \mbox{Erf} \left[ \sqrt{\frac{m_q^2}{8 \beta^2 x \left(1-x \right)}}
\right]\right). \nonumber \\
\label{eq.WH}
\end{eqnarray}
}
The pion distribution amplitude at the initial scale is controlled by 
the parameter B. They have found that the BABAR data for pion transition
form factor at low and high transferred four-momentum squared regions 
can be described by setting $B \approx$  0.6. 
This pion distribution amplitude is rather similar to the well know 
Chernyak-Zhitnitsky \cite{CZ} distribution amplitude 
($\phi_{\pi \, CZ} = 30x(1-x)(2x-1)^2$).
In the following we shall use $B=$ 0.6 and $m_q=$ 0.3 GeV. 
Then $A=$ 16.62 GeV$^{-1}$ and $\beta=$ 0.745 GeV. 

The total (angle integrated) cross section 
for the process can be expressed in terms 
of the amplitude of the process discussed above as:
\begin{equation}
\sigma _{\gamma \gamma \to \pi \pi } = 
\int \frac{2 \pi}{4 \cdot 64 \pi^2 W^2 } \frac{p}{q} 
\sum_{\lambda_1, \lambda_2} 
\left|  \mathcal{M}  \left( \lambda_1, \lambda_2 \right) \right|^2 dz \;,
\end{equation}
where the factor $4$ is due to averaging over initial photon helicities.

The hand-bag model was proposed as an alternative
for the leading term Brodsky-Lepage pQCD approach \cite{HB}.
It is based on the philosophy that the present
energies are not sufficient for the dominance 
of the leading pQCD terms. As in the case of 
BL pQCD the hand-bag approach applies at large
Mandelstam variables $s \sim -t \sim -u$ i.e. at large 
momentum transfers. Diehl, Kroll and Vogt presented 
a sketchy derivation \cite{HB} obtaining that
the angular dependence of the amplitude is 
$ \propto 1/ \sin^2 \theta$. 
In this approach the ratio of the cross section for the $\pi^0 \pi^0$
process to that for the $\pi^+ \pi^-$ process does not depend on $\theta$ 
and is $\frac{1}{2}$. The nonperturbative object 
$R_{\pi \pi} \left(s \right)$ in the hand-bag amplitude,
describing transition from a quark pair to a meson pair, 
cannot be calulated from first principles.
In Ref. \cite{HB} the form factor was parametrized in terms 
of the valence and non-valence form factors as:
\begin{equation}
R_{\pi \pi} \left(s \right) = 
\frac{5}{9s} a_u \left( \frac{s_0}{s} \right)^{n_u} + 
\frac{1}{9s} a_s \left( \frac{s_0}{s} \right)^{n_s}.
\end{equation}
The $a_u$, $n_u$, $a_s$ and $n_s$ values 
found from the fit in Ref. \cite{HB}
slightly depend on energy.
For simplicity we have averaged these values 
and used:
$a_u=$ 1.375 GeV$^2$, $n_u=$ 0.4175, 
$a_s=$ 0.5025 GeV$^2$ and $n_s=$ 1.195.

In Ref.\cite{KS_pipi} we have discussed in detail elementary cross
sections as a function of photon-photon energy and as a function
of $\cos(\theta)$. Here we will present only nuclear cross 
sections calculated within EPA discussed in the theoretical section.

\begin{figure}[htb]                  
\begin{center}
\includegraphics[width=0.35\textwidth]{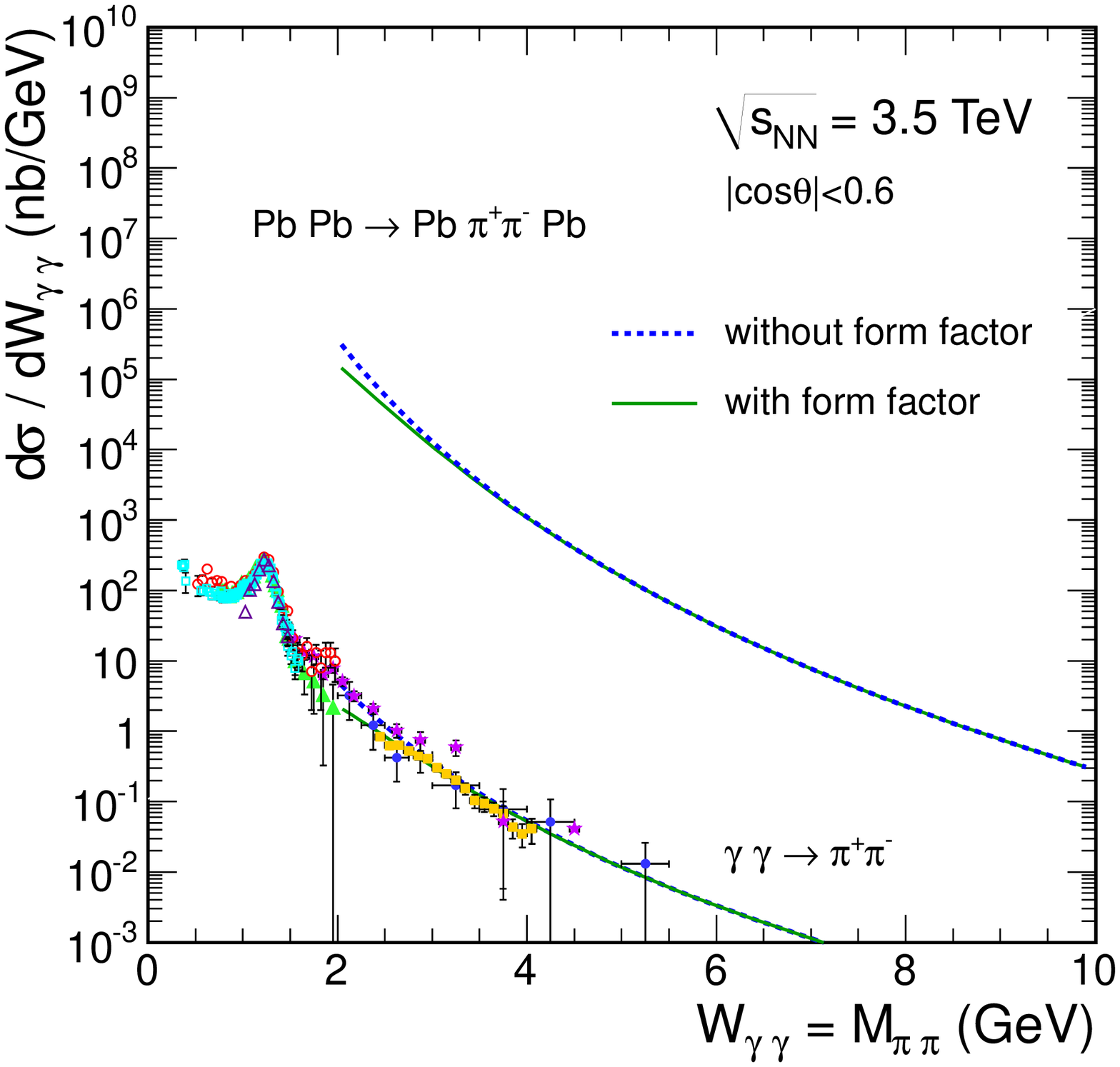}
\includegraphics[width=0.35\textwidth]{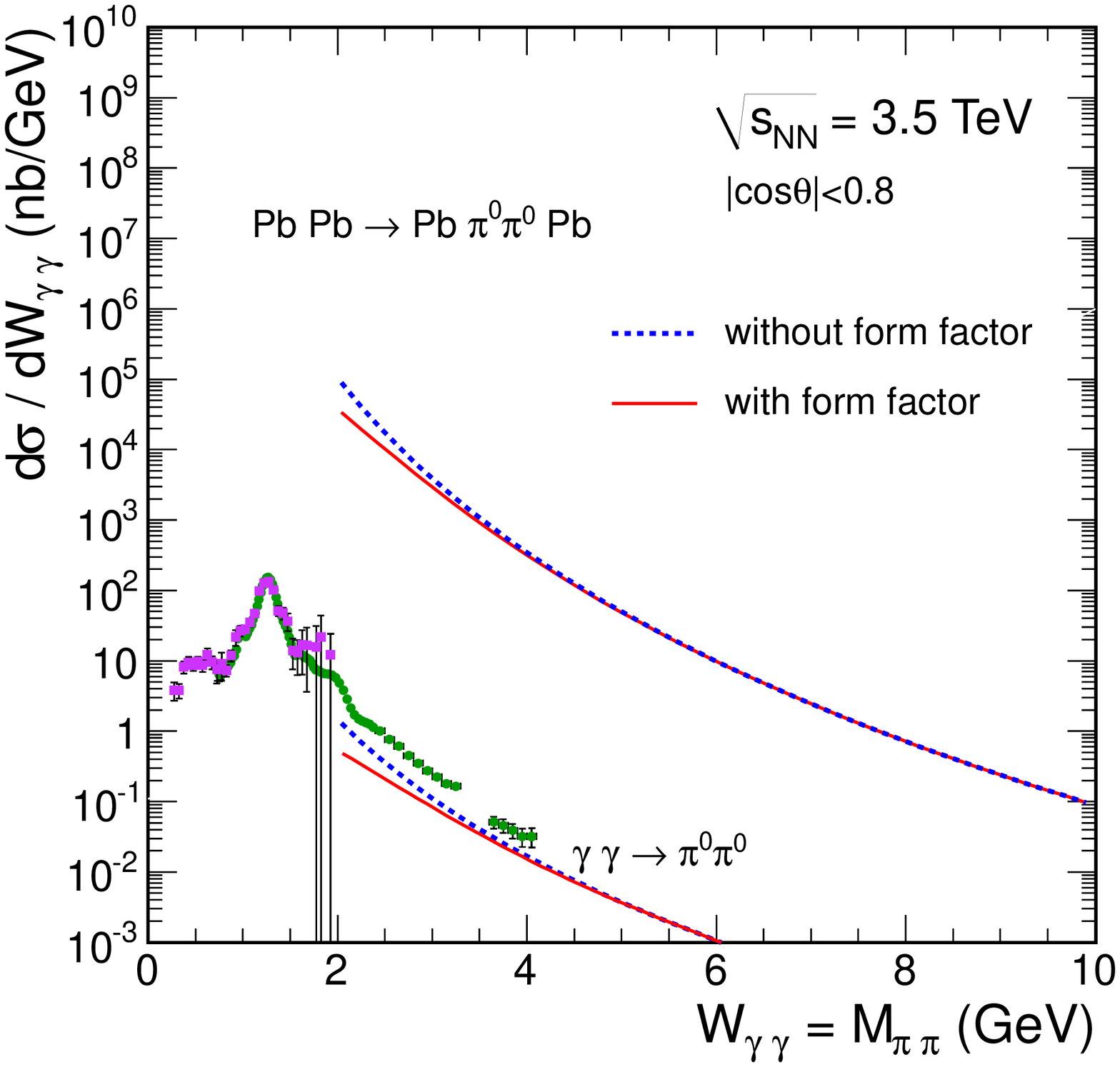}
\end{center}
   \caption{\label{fig:dsig_dw}\textsl{}
   \small The nuclear (upper lines) and elementary (lower lines) cross section
   as a function of photon--photon subsystem energy 
   $W_{\gamma \gamma}$ in the b-space EPA within the BL pQCD approach
   for the elementary cross section with Wu-Huang distribution amplitude. 
   The angular ranges in the figure caption
   correspond to experimental cuts.
}
\end{figure}

In Fig. \ref{fig:dsig_dw} we show distribution in the two-pion invariant mass
which by the energy conservation is also the photon-photon subsystem energy.
For this figure we have taken experimental limitations usually used
for the $\pi \pi$ production in $e^+ e^-$ collisions. 
In the same figure we show our results for the $\gamma \gamma$ collisions 
extracted from the $e^+ e^-$ collisions together with the corresponding 
nuclear cross sections for $\pi^+ \pi^-$ (left panel)
and $\pi^0 \pi^0$ (right panel) production.
We show the results for the standard BL pQCD approach with and without
extra form factor (see \cite{SS2003}).


Comparing the elementary and nuclear cross sections
we see a large enhancement of the order of 10$^4$ which is, however, somewhat less
than $Z_1^2 Z_2^2$ one could expect from a naive counting.
%

\subsection{Exclusive production of $\rho^0 \rho^0$ pairs}

\begin{figure}[htb]
\begin{center}
\includegraphics[width=0.35\textwidth]{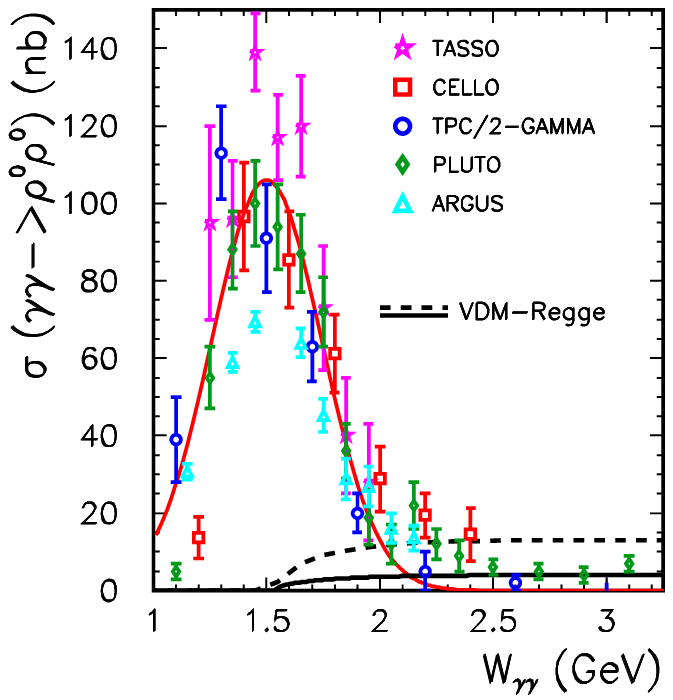}
\includegraphics[width=0.35\textwidth]{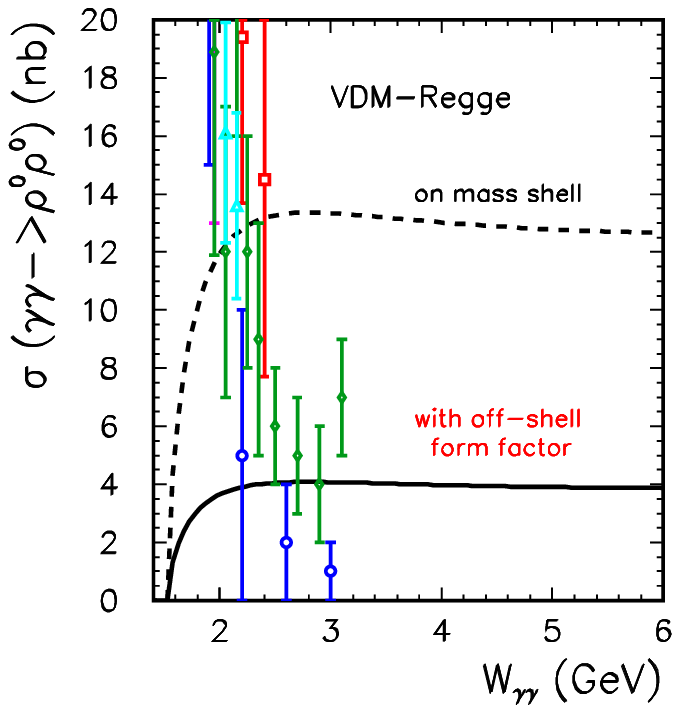}
\end{center}
\caption{
\label{fig:gg_rr_cs}\textsl{}
\small Elementary cross section for $\gamma \gamma \to \rho^0 \rho^0$
reaction. The fit to the experimental data is shown in the left panel
and our predictions for the high energy in the right panel.}
\end{figure}		

At low energies one observes a huge enhancement of the cross section
for the elementary process $\gamma \gamma \to \rho^0 \rho^0$ (see left
panel of Fig.\ref{fig:gg_rr_cs}). In the right panel we show predictions
of a simple Regge-VDM model with parameters adjusted to the world
hadronic data.
More details about our model can be found in our
original paper \cite{KSS_rhorho}.

\begin{figure}[htb]
\begin{center}
   \includegraphics[width=0.35\textwidth]{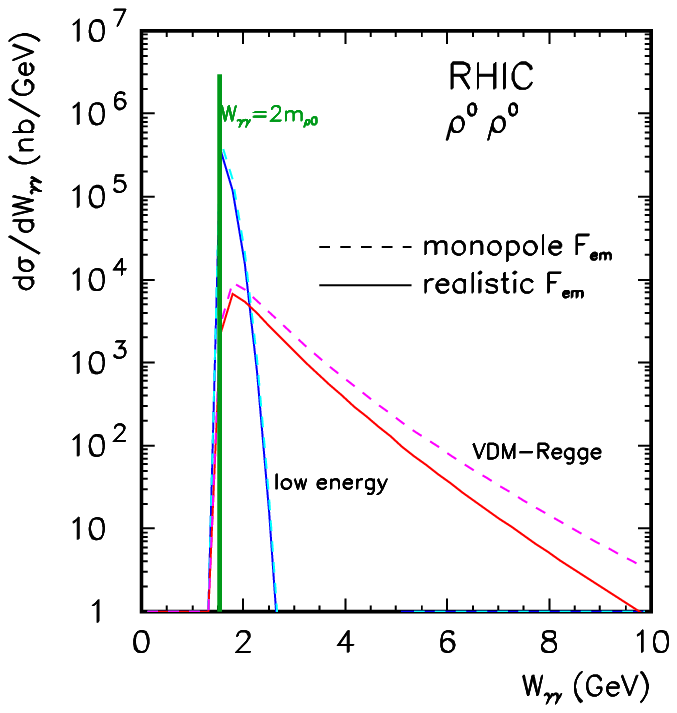}
   \includegraphics[width=0.35\textwidth]{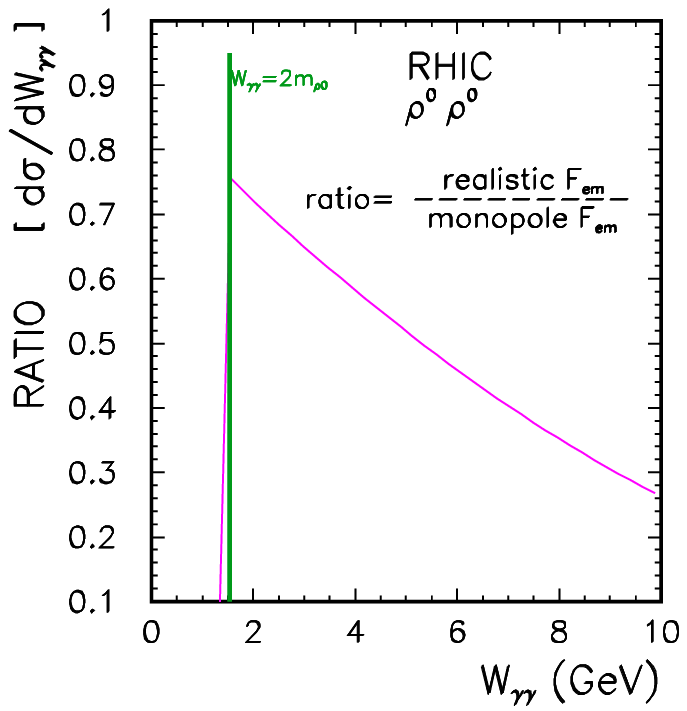}
\end{center}
\caption{Distribution in the $\rho^0 \rho^0$ invariant mass}
\label{fig:dsig_dw_rho0rho0}
\end{figure}

In Fig.\ref{fig:dsig_dw_rho0rho0} we show distribution in $\rho^0\rho^0$
invariant mass (left panel) and the ratio of the cross section for
realistic and monopole form factors.

\subsection{Some comments and outlook}

We have presented some examples of processes that could be soon studied
at RHIC or LHC. In all cases we have obtained measurable cross sections.
We have pointed out that the inclusion of realistic charge form factor
is necessary to obtain realistic particle distributions.

Measurements of the processes discussed here are not easy as one has to 
assure exclusivity of the process, i.e., it must be checked that there are no
other particles than that measured in central detectors.
In all cases fissibilty studies, including Monte Carlo simulations, are 
required. 

In the close future one may expect results for two-pion
and single vector mesons production from LHC experiments. 
Exclusive production of one or two pairs of charged leptons should be 
feasible too.

\vspace{0.3cm}

{\bf Acknowledgments}
Some of the results presented here were obtained in collaboration with 
Wolfgang Sch\"afer, Valerij Serbo and Magno Machado.

\vspace{0.3cm}

\section*{References}



\begin{thebibliography}{99}

\bibitem{review}
Budnev V M et al. 1975
{\it Phys. Rep.} {\bf 15} 4\\
Bertulani C A, Baur G 1988 {\it Phys. Rep.} {\bf 163} 299\\
Baur G, Hencken K, Trautmann D, Sadovsky S and Kharlov Y 2002
{\it Phys. Rep.} {\bf 364} 259\\
Bertulani C A, Klein S and Nystrand J 2005
{\it J Ann. Rev. Nucl. Part. Sci.}{\bf 55} 271\\
Baltz A et al. 2008, {\it Phys. Rep.} {\bf 458} 1

\bibitem{KS_mumu}
K{\l}usek-Gawenda M and Szczurek A 2010
{\it Phys. Rev.} C {\bf 82} 014904

\bibitem{KSMS_qqbar}
K{\l}usek-Gawenda M, Szczurek A, Machado M and Serbo V 2011
{\it Phys. Rev.} C {\bf 83} 024903

\bibitem{KSS_rhorho}
K{\l}usek M, Sch\"afer W and Szczurek A 2009
{\it Phys. Lett.} B {\bf 674} 92 

\bibitem{KS_pipi}
K{\l}usek-Gawenda M and Szczurek A 2011, 
{\it Phys. Lett.} B {\bf 700} 322

\bibitem{LS_DDbar}
{\L}uszczak M and Szczurek A 2009
{\it Phys. Lett.} B {\bf 700} 116

\bibitem{Jackson}
Jackson J D 1975, {\it Classical Electrodynamics, 2nd ed.} (Wiley,
New York), p. 722

\bibitem{HTB94}
Hencken K, Trautmann D and Baur G 1994
{\it Phys. Rev.} A {\bf 49} 1584

\bibitem{Barrett_Jackson}
Barrett R C and Jackson D F 1977 {\it Nuclear Sizes and Structure}, 
(Clarendon Press, Oxford)

\bibitem{nuclear_density}
de Vries H, de Jager C W and de Vries C 1987
{\it Atomic Data and Nuclear Data Tables} {\bf 36} 495

\bibitem{BB87}
Baur G and Bertulani C A 1987
{\it Phys. Rev.} C {\bf 35} 836

\bibitem{HTB99}
Hencken K, Trautmann D and Baur G 1999
{\it Phys. Rev.} C {\bf 59} 841

\bibitem{Serbo}
Ivanov D Yu, Schiller A and Serbo V G 1999
{\it Phys. Lett.} B {\bf 454} 155\\
Henecken K, Kuraev E A and Serbo V G 2007
{\it Phys. Rev.} C {\bf 75} 034903\\
Jentschura U D, Hencken K and Serbo V G 2008
{\it Eur. Phys. J.} C {\bf 58} 281\\
Jentschura U D and Serbo V G 2009
{\it Eur. Phys. J.} C {\bf 64} 309

\bibitem{Baltz}
Baltz A 2008 {\it Phys. Rev. Lett.} {\bf 100} 062302\\
Baltz A 2009 {\it Phys. Rev.} C {\bf 80} 034901

\bibitem{KKMV09}
Kniehl B A, Kotikov A V, Merebashvili Z V and Veretin O L 2009
{\it Phys. Rev.} D {\bf 79} 114032

\bibitem{TKM}
Timneanu N, Kwieci\'nski J and Motyka L 2002
{\it Eur. Phys. J.} C {\bf 23} 513

\bibitem{szczurek}
Szczurek A 2002
{\it Eur. Phys. J.} C {\bf 26} 183

\bibitem{GVR1992}
Gluck M, Reya E and Vogt A 1992
{\it Phys. Rev.} D {\bf 46} 1973

\bibitem{BL81}
Brodsky S J and Lepage G P 1981
{\it Phys. Rev.} D {\bf 24} 1808

\bibitem{JA}
Ji C R and Amiri F 1990
{\it Phys. Rev.} D {\bf 42} 3764

\bibitem{BL79}
Brodsky S J and Lepage G P 1979
{\it Phys. Lett.} B {\bf 87} 359

\bibitem{ER}
Efremov A V and Radyushkin A V 1980
{\it Phys. Lett.} B {\bf 94} 245

\bibitem{WH}
Wu X G and Huang T 2010
{\it Phys. Rev.} D {\bf 82} 034024

\bibitem{CZ}
Chernyak V L and Zhitnitsky A R 1982
{\it Nucl. Phys.} B {\bf 201} 492

\bibitem{HB}
Diehl M, Kroll P and Vogt C 2002
{\it Phys. Lett.} B {\bf 532} 99\\
Diehl M and Kroll P 2010
{\it Phys. Lett.} B {\bf 683} 165

\bibitem{SS2003}
Szczurek A and Speth J 2003
{\it Eur. Phys. J.} A {\bf 18} 445

\end{thebibliography}
\end{document}